\documentclass{article}
\usepackage{spconf,amsmath,graphicx}
\usepackage{cite,bm}
\usepackage{amssymb}
\usepackage{amsfonts}
\usepackage{comment}
\usepackage{amsmath,mathtools}
\usepackage{booktabs}
\AtBeginDocument{
  \abovedisplayskip     =0.4\abovedisplayskip
  \abovedisplayshortskip=0.4\abovedisplayshortskip
  \belowdisplayskip     =0.4\belowdisplayskip
  \belowdisplayshortskip=0.4\belowdisplayshortskip}

\title{REMIX-CYCLE-CONSISTENT LEARNING ON ADVERSARIALLY LEARNED SEPARATOR FOR ACCURATE AND STABLE UNSUPERVISED SPEECH SEPARATION}

\name{
    Kohei Saijo,
    Tetsuji Ogawa
}
    
\address{
Department of Communications and Computer Engineering, Waseda University, Tokyo, Japan
}

\begin{document}
\maketitle
\begin{abstract}
A new learning algorithm for speech separation networks is designed to explicitly reduce residual noise and artifacts in the separated signal in an unsupervised manner.
Generative adversarial networks are known to be effective in constructing separation networks when the ground truth for the observed signal is inaccessible.
Still, weak objectives aimed at distribution-to-distribution mapping make the learning unstable and limit their performance.
This study introduces the remix-cycle-consistency loss as a more appropriate objective function and uses it to fine-tune adversarially learned source separation models.
The remix-cycle-consistency loss is defined as the difference between the mixed speech observed at microphones and the pseudo-mixed speech obtained by alternating the process of separating the mixed sound and remixing its outputs with another combination.
The minimization of this loss leads to an explicit reduction in the distortions in the output of the separation network.
Experimental comparisons with multichannel speech separation demonstrated that the proposed method achieved high separation accuracy and learning stability comparable to supervised learning. \end{abstract}


\section{Introduction}
\label{sec:intro}

Since speech recordings are frequently contaminated by interference and background noise, source separation technology has become essential in voice applications.
Deep neural networks (DNNs) have been shown to give remarkably high separation performance~\cite{dc,danet,pit}, especially in supervised settings where models are trained using large number of pairs of observed signals and corresponding ground truths (i.e., distortion-free signals).
It, however, is not feasible to collect such distortion-free data in real environment, and learning source separation models under unsupervised conditions where paired data are not available is thus highly desired.

Adversarial learning~\cite{gan} of source separation models aims to bring the distribution of separated signals closer to the distribution of distortion-free signals.
It was originally introduced in supervised settings~\cite{segan,mmsegan,metricgan,hifigan,tassan,improved_segan} but is more potent in unsupervised situations where the ground truth is not given~\cite{remix,unpaired_ssgan,unpaired_segan,ccse1,ccse2,ccse3}.
However, its weak objectives aimed at distribution-to-distribution mapping destabilize the learning (i.e., performance is highly dependent on initial parameters) and limit the separation performance.
Several attempts have been made to overcome these shortcomings by exploiting the remix-cycle-consistency loss~\cite{remix} and the boundary equilibrium generative adversarial network~\cite{unpaired_ssgan}.

In contrast, this study attempts to solve the above problem by imposing a more direct requirement that reduces residual noise and artifacts in the separated signal on unsupervised learning of source separation models.
For this purpose, a two-stage learning algorithm is proposed: the neural separator (specifically, a minimum variance distortionless response (MVDR) beamformer~\cite{mvdr} with DNN-based time-frequency (TF) masking \cite{dnn_mvdr_hakan,dnn_mvdr_heymann,dnn_mvdr_ochiai}) is adversarially trained and then fine-tuned using remix-cycle-consistency loss.
This loss is defined as the difference between the mixed sound observed at microphones and the pseudo-mixed sound obtained by alternating the process of separating the mixed sound and remixing its outputs with another combination.
Since the repetitive process of unmixing and remixing accumulates distortions, minimizing the remix-cycle-consistency loss leads to the removal of residual noise and artifacts in the separator's output.
In this case, using the observed mixture as a teacher means imposing constraints on learning at the input speech level rather than at the distribution level.
Therefore, tuning the separator to minimize this loss is expected to improve the separation accuracy and reduce the instability of adversarial learning.

There are two key contributions made in this work: 
1) we provide an algorithm that simultaneously addresses the performance limitation and learning instability of neural separators adversarially learned in an unsupervised manner, and 2) we provide insights into the effects of using remix-cycle-consistent learning in the latter stage of two-stage learning: this would be appropriate to be used for fine-tuning of the neural separators.

The remainder of this paper is organized as follows.
Section \ref{sec:related_work} provides the elemental technologies used.
Section \ref{sec:proposed_method} describes the details of the proposed two-stage learning algorithm for stable and accurate speech separation.
Section \ref{sec:experiments} demonstrates the effectiveness of the proposed method through experiments.
Section \ref{sec:conclusion} concludes this paper.

\section{Elemental Technology}
\label{sec:related_work}

Let us denote mixtures of $N$ sources captured by $M$ microphones in the short-time Fourier transform (STFT) domain as $\bm{x}_{f,t}\in\mathbb{C}^{M}$,
where $f$ and $t$ denote a frequency bin and a time index, respectively, and are omitted hereafter for simplicity.

\subsection{Adversarial Learning}
\label{ssec:gan}

GANs~\cite{gan} consist of two modules, a generator and a discriminator.
The discriminator tries to distinguish real data from the generator's output (i.e., fake data),
while the generator tries to yield data that can fool the discriminator,
which leads to the generator that outputs data similar to real data.
Optimization of GANs is formulated as:
\begin{align}
  \label{eqn:gan_cost}
    \min_{G}\max_{D}\mathcal{L}_{\rm GAN}=
    \mathbb{E}_{\bm{y}\sim{p_{\rm data}(\bm{y})}}[\log{D(\bm{y})}] \nonumber \\
    + \mathbb{E}_{\bm{x}\sim{p_{\bm{x}}(\bm{x})}}[\log{(1-D(G(\bm{x})))}],
\end{align}
where $G$ and $D$ denote the generator and the discriminator, and $\bm{y}$ and $\bm{x}$ represent the real and the fake, respectively.
This equation implies that
the generator is trained to make the distribution of its output closer to the real data distribution.

When applying GANs to speech separation,
the generator is the separator to be sought, and the real data are distortion-free (clean) speech signals.
GANs were originally introduced for speech separation in a supervised setting where the ground truth for each observed mixture could be given~\cite{segan}.
This framework, however, is more potent in unsupervised settings~\cite{unpaired_segan,unpaired_ssgan} where any clean signal that is not paired with an observed mixture can be used as real data,
because its objective is based on distribution-to-distribution mapping.

\subsection{Adversarial Unmix-and-Remix}
\label{ssec:unmix_and_remix}

An attempt was made to exploit cycle-consistent learning for single-channel unsupervised speech separation~\cite{remix}.
The cycle-consistency loss introduced in this study has been utilized in our proposal.

Let $\bm{x}_{j}$ be a mixture of two sound sources,
$s_1^{(\bm{x}_j)}$ and  $s_2^{(\bm{x}_j)}$.
Suppose that $\bm{x}_{1}$ and $\bm{x}_{2}$ are given as mixtures of different source pairs,
$s_1^{(\bm{x}_1)}$ and  $s_2^{(\bm{x}_1)}$, and $s_1^{(\bm{x}_2)}$ and  $s_2^{(\bm{x}_2)}$,
which are all assumed to be different speech sources.
First, these mixtures are segregated by the separator $G$ as follows:
\begin{align}
  \label{eqn:separation}
    \bm{\hat{s}}_{1}^{(\bm{x}_1)}, \bm{\hat{s}}_{2}^{(\bm{x}_1)} = G(\bm{x}_1), \quad
    \bm{\hat{s}}_{1}^{(\bm{x}_2)}, \bm{\hat{s}}_{2}^{(\bm{x}_2)} = G(\bm{x}_2),
\end{align}
where $\bm{\hat{s}}_{i}^{(\bm{x}_{j})}$ denotes the $i$-th separated signal
of the $j$-th observed mixture $\bm{x}_{j}$.
These outputs are remixed in combinations between different observations as follows:
\begin{align}
  \label{eqn:remix}
    \bm{z}_1 = \bm{\hat{s}}_{1}^{(\bm{x}_1)} + \bm{\hat{s}}_{2}^{(\bm{x}_2)}, \quad
    \bm{z}_2 = \bm{\hat{s}}_{1}^{(\bm{x}_2)} + \bm{\hat{s}}_{2}^{(\bm{x}_1)}.
\end{align}
If speech separation in (\ref{eqn:separation}) works perfectly, the distribution of such {\it{pseudo mixtures}} will be consistent with that of the observed mixtures.
$G$, therefore, can be trained adversarially with the discriminator $D$ that aims to discriminate between the observed mixtures (real) and the pseudo mixtures (fake).
In addition to the loss function of this framework,
$\mathcal{L}_{\rm GAN}$,
two additional auxiliary loss functions are introduced as follows.

The pseudo mixtures $\bm{z}_1$ and $\bm{z}_2$ are unmixed by the same separator $G$ as follows:
\begin{align}
  \label{eqn:remix_separation}
    \bm{\hat{s}}_{1}^{(\bm{z}_1)}, \bm{\hat{s}}_{2}^{(\bm{z}_1)} = G(\bm{z}_1), \quad
    \bm{\hat{s}}_{1}^{(\bm{z}_2)}, \bm{\hat{s}}_{2}^{(\bm{z}_2)} = G(\bm{z}_2).
\end{align}
These outputs
are remixed in the closest combination to the observed mixtures $\bm{x}_j$ into another pseudo mixtures as:
\begin{align}
  \label{eqn:mix_perm}
    \bm{\hat{x}}_{j} = \left[\bm{A}\bm{\hat{S}}^{(\bm{z})}\right]_{j}, \quad \bm{A} = \min_{\bm{A}} \left\lVert \bm{x}_{j} - \left[\bm{A}\bm{\hat{S}}^{(\bm{z})}\right]_{j} \right\rVert,
\end{align}
where $\bm{\hat{S}}^{(\bm{z})} \!=\! \Big[
{\bm{\hat{s}}_{1}^{(\bm{z}_1)}}, {\bm{\hat{s}}_{2}^{(\bm{z}_1)}}, {\bm{\hat{s}}_{1}^{(\bm{z}_2)}}, {\bm{\hat{s}}_{2}^{(\bm{z}_2)}}
\Big]^\top{} \!\in\! \mathbb{C}^{2N \!\times\! M}$ denotes a set of separated signals of $\bm{z}_1$ and $\bm{z}_2$,
and $\bm{A} \!\in\! \mathbb{B}^{2 \!\times\! 2N}$ denotes a binary matrix
that represents the assignment of each separated signal to $\bm{x}_{1}$ or $\bm{x}_{2}$.
The cycle-consistency loss (also referred to as remix-cycle-consistency loss in this paper) is defined as:
\begin{align}
  \label{eqn:closs}
    \mathcal{L}_{C} = \sum\nolimits_{j} || \bm{x}_j - \bm{\hat{x}}_{j} ||.
\end{align}
Since $\bm{\hat{s}}_i^{(\bm{x}_j)}$ contains residual noise and artifacts,
the pseudo mixtures,
$\bm{z}_j$, will contain unobserved distortions,
which can make $\bm{\hat{s}}_i^{(\bm{z}_j)}$ distorted.
It implies that 
distortion reduction in $\bm{\hat{s}}_i^{(\bm{x}_j)}$ leads to that in $\bm{\hat{s}}_i^{(\bm{z}_j)}$.
Therefore, minimizing the loss $\mathcal{L}_C$,
which is computed using $\bm{\hat{s}}_i^{(\bm{z}_j)}$, 
facilitates reducing the residual noise and artifacts in $\bm{\hat{s}}_i^{(\bm{x}_j)}$.

Here, it should be noted that
a trivial solution $\bm{\hat{s}}_{1}^{(\bm{x}_j)} \!=\! \bm{x}_j$ and $\bm{\hat{s}}_{2}^{(\bm{x}_j)} \!=\! \bm{0}$ is unduly optimal
both in $\mathcal{L}_{\rm GAN}$
and in $\mathcal{L}_{C}$.
To compensate for this deficiency and to facilitate the separation,
the following loss function is also introduced:
\begin{align}
  \label{eqn:eloss}
    \mathcal{L}_{E} = \sum\nolimits_{j} {\left(\left(\bm{\hat{s}}_{1}^{(\bm{x}_j)}\right) ^2 + \left(\bm{\hat{s}}_{2}^{(\bm{x}_j)}\right)^2\right)}.
\end{align}

$\mathcal{L}_{C}$ is designed to reduce distortions in the separated sound explicitly, but $\mathcal{L}_{\rm GAN}$ does not because it can be small even if $\bm{\hat{s}}_i^{(\bm{x}_j)}$ contains distortions depending on the combinations that make up $\bm{z}_j$.
In fact, this method performed well in image separation but did not work in speech separation~\cite{remix}.


\section{Remix-cycle-consistent learning on adversarially-learned separator}
\label{sec:proposed_method}

\begin{figure*}[t]
\centering
\centerline{\includegraphics[width=0.95\linewidth]{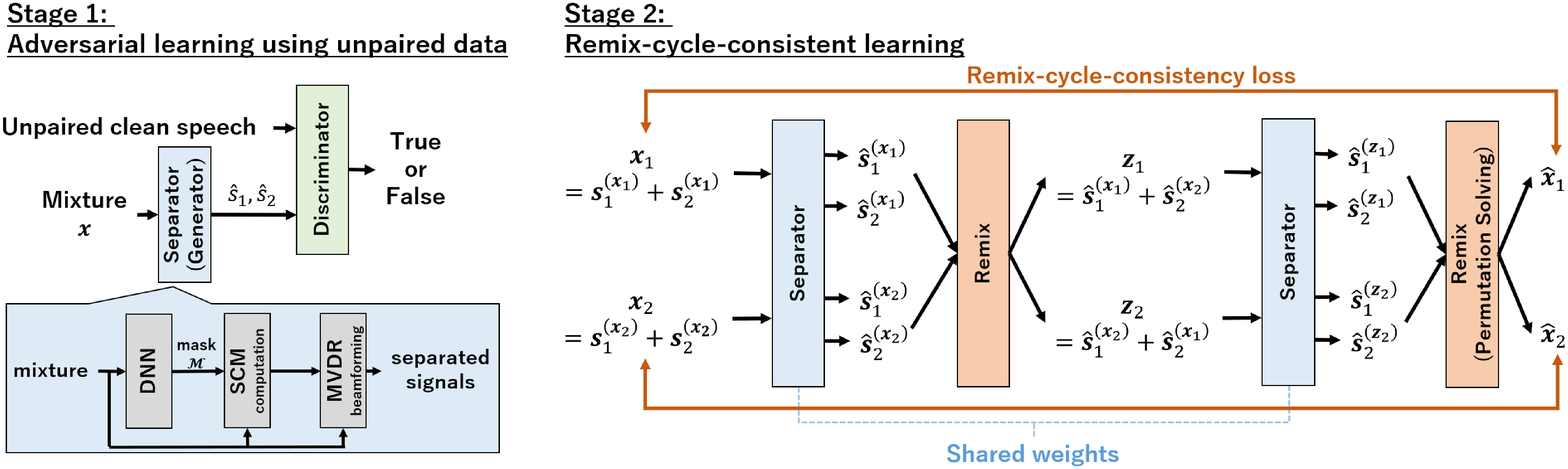}}
\vspace{-2mm}
\caption{Overview of proposed method.
Neural separator (DNN-based MVDR beamformer) is adversarially trained
and then tuned using remix-cycle-consistent learning.
$\bm{x}_1$ and $\bm{x}_2$ are different observed mixtures (not observed simultaneously). }
\label{fig:overview}
\vspace{-2mm}
\end{figure*}

This section describes an algorithm for training a separation network
to explicitly remove residual noise and artifacts
in the separated signals in the unsupervised setting. 
The overview of this method is illustrated in Fig.~\ref{fig:overview}.
The neural separator is adversarially trained to output a distortion-free speech-like signal coarsely and then tuned using a remix-cycle-consistency loss described in (\ref{eqn:closs})
to reduce distortions at the separator's output finely for each utterance.
Note that $\bm{x}_1$ and $\bm{x}_2$ in Fig.~\ref{fig:overview} are different observed mixtures (i.e., they are not other channels of sound observed simultaneously).

\subsection{Neural Separator}

The separator employs a DNN-based MVDR beamformer \cite{dnn_mvdr_hakan,dnn_mvdr_heymann,dnn_mvdr_ochiai},
as illustrated at the bottom left of Fig.~\ref{fig:overview}.
The DNN takes a mixture as an input and estimates a TF mask of each source $\mathcal{M}_i$.
The spatial covariance matrice (SCM) for the speech  $\bm{R}_{i}^{(s)}$
and that for the noise $\bm{R}_{i}^{(n)} \!\in\! \mathbb{C}^{M \times M}$
are then computed using $\mathcal{M}_i$ and $\bm{1}-\mathcal{M}_i$,
respectively~\cite{mask_scm}.
The filter coefficient of the MVDR beamformer~\cite{mvdr2} is computed as:
\begin{align}
    \label{eqn:mvdr}
    {\bm{W}}_{i} = \frac{{\bm{R}_{i}^{(n)}}{\bm{R}_{i}^{(s)}}}{{\rm tr}({\bm{R}_{i}^{(n)}}{\bm{R}_{i}^{(s)}})} \in \mathbb{C}^{M \times M}.
\end{align}
The $i$-th separated signal of a mixture $\bm{x}$ is given as:
\begin{align}
    \label{eqn:filtering}
    \bm{\hat{s}}_{i}^{(\bm{x})} = {\bm{W}}_{i}^{\mathsf{H}}{\bm{x}} \in\mathbb{C}^{M}.
\end{align}

The pseudo mixtures need to be simulated as observed at the microphone locations to calculate the remix-cycle-consistency loss with the observed mixtures.
To do this, the separator's output (i.e., estimated source) will be multiplied by the steering vector for the $M$-channel microphone positions.
Note that (\ref{eqn:mvdr}) is formulated with such a process, and there is no need to apply the steering vectors in (\ref{eqn:filtering}).

\subsection{Learning Algorithm of Neural Separator}

The neural separator, which corresponds to a generator, is adversarially trained with a discriminator that distinguishes the separator's outputs from unpaired clean speech.
Since adversarial learning only aims to learn a distribution-to-distribution mapping,
the performance of the separator trained with such a scheme can be limited and highly dependent on the initial values of the parameters.

The neural separator coarsely trained in the previous stage is then finely tuned 
using the remix-cycle-consistency loss
introduced in \cite{remix}.
As described in Sect.~\ref{ssec:unmix_and_remix},
this loss is designed to explicitly reduce the distortions in the separated signal for each utterance.
Such a per-input-sample requirement is imposed on the learning instead of coarser constraints at the distribution level as in GANs, which can improve the separation performance and stabilize the learning.

The effectiveness of the proposed method will be further discussed in terms of 1) design for optimization and 2) design for loss functions.
First, two-stage learning used in the proposed method is essential for improving the separation accuracy and stabilizing the learning.
The remix-cycle-consistency loss may interfere with learning when the separator's performance is low because the trivial solution is optimal.
This loss, therefore, should be used for tuning a well-trained separator,
as in the proposed method.
In contrast, the existing method~\cite{remix} that minimizes all losses simultaneously did not work.
Second, the adversarial loss in the proposed method is defined on the speech domain, i.e., to explicitly learn mixture-to-clean mapping, which is the aim of speech separation. In contrast, the adversarial loss in the existing method is defined on the mixture domain.
In addition,
the proposed method does not need the separation-promoting loss described in (\ref{eqn:eloss}), unlike the existing method.
The well-designed (i.e., not unduly flexible) neural separator, as well as two-stage learning, may help the proposed method avoid being trapped in the trivial solution.







\begin{figure*}[t]
\centering
\centerline{\includegraphics[width=0.95\linewidth]{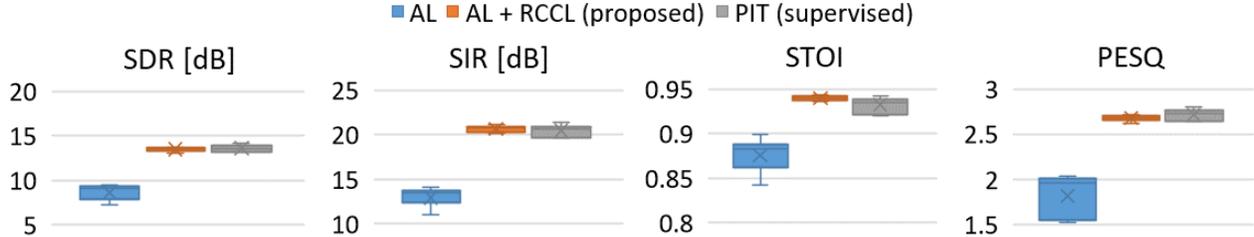}}
\vspace{-2mm}
\caption{Stability of adversarial learning (AL), AL followed by remix-cycle-consistent learning (AL+RCCL), and permutation invariant training (PIT), represented by box plots of SDR, SIR, STOI, and PESQ in ten trials (with different seeds) for learning.}
\label{fig:result_box}
\vspace{-2mm}
\end{figure*}

\section{Speech separation experiment}
\label{sec:experiments}

Experimental comparisons were conducted to verify the effectiveness of the proposed method on the accuracy of speech separation and the stability of learning.
To this end, the following learning methods were compared:
\begin{itemize}
\vspace{-1mm}
\setlength{\itemsep}{0pt}
\setlength{\parskip}{0pt}
\setlength{\leftskip}{-10pt}

\item {\bf AL:} adversarial learning using unpaired clean speech (only the first stage in the proposed method)

\item {\bf AL + RCCL:} adversarial learning followed by remix-cycle-consistent learning (proposed method)

\item {\bf PIT:} permutation invariant training (upper limit)

\vspace{-1mm}
\end{itemize}
Note that PIT~\cite{pit} is a supervised learning method and will give an upper limit on AL and AL+RCCL, which are unsupervised learning methods.
In addition, the adversarial unmix-and-remix is not included in the comparison because it was reported that it did not work in speech separation~\cite{remix}.

\begin{table}[t]
\vspace{-2mm}
\begin{center}
\caption{Speech separation performance of neural separator trained with adversarial learning (AL), AL followed by remix-cycle-consistent learning (RCCL), and permutation invariant training (PIT) in terms of SDR, SIR in decibels, STOI, and PESQ averaged over ten trials.}
\label{table:result}
\vspace{3mm}
\begin{tabular*}{\linewidth}{@{}l@{\extracolsep{\fill}}cccc@{}}
\toprule
{Method}  &{SDR} &{SIR} &{STOI} &{PESQ} \\
\midrule
Observation & 0.00  &0.175  &0.712  &1.16   \\
\midrule
USV: AL &8.58  &12.9  &0.875  &1.81   \\
USV: AL + RCCL&\bf{13.4}  &\bf{20.6}  &\bf{0.939}  &\bf{2.68}   \\
\midrule
SV: PIT \cite{pit}   &13.6  &20.4  &0.932  &2.72   \\
\bottomrule

\end{tabular*}
\end{center}
\vspace{-4mm}
\end{table}

\subsection{Speech Materials}
\label{sec:setup}

An anechoic sound field with two sources observed by four microphones whose spacings are 3~cm was simulated using pyroomacoustics~\cite{pra}.
The distance between the sound sources and the center of the microphone array was 1~m.
The directions of two sound sources were randomly chosen without duplication between -90$^{\circ}$ and 90$^{\circ}$ at 15$^{\circ}$ intervals (0$^{\circ}$ is the front).
Speech sources were sampled at 16~kHz and selected from the Wall Street Journal (WSJ0) corpus~\cite{wsj0}.
10240, 1024, and 512 microphone observations (mixtures) were used for training, validation, and testing, respectively.
Another 10240 distortion-free utterances were used as unpaired clean speech for adversarial learning.
Speech signals were analyzed using the Hanning window with an FFT length of 512 and a hop size of 128.

\subsection{Network Architecture}
\label{sec:architecture}

Neural networks were employed in the generator $G$
(specifically, the mask estimator in the neural separator)
and the discriminator $D$.
The mask estimator is composed of a fully connected layer with a rectified linear unit activation
and two layers of bi-directional long short-term memory with 500 units
followed by another fully connected layer with the softmax activation.
$D$ consists of four two-dimensional convolutional layers
followed by the sigmoid activation.
All the parameters were optimized using the Adam optimizer~\cite{adam}.
In the first stage, the learning rate was set to $5.0 \times 10^{-4}$, and the batch size of $G$ and that of $D$ were set to 32 and 64, respectively.
Note that each mixture consists of two speech sources, so the batch size of $D$ was twice that of $G$.
In the second stage, the same learning rate was used, but the batch size was 16.

\subsection{Experimental Results}
\label{sec:result}

The output signals of the neural separators
trained with three learning methods and the microphone observation were evaluated
using four criteria:
the signal-to-distortion ratio (SDR),
the signal-to-interference ratio (SIR),
the short-time objective intelligibility (STOI)~\cite{stoi},
and the perceptual evaluation of speech quality (PESQ)~\cite{pesq}.
The average values of the performance over ten trials
(i.e., the models created with ten different seeds) are listed in Table \ref{table:result}.
The variation in performance, which represents the dependence on the initial values, is shown in Fig.~\ref{fig:result_box} as box plots, where a five-number summary was the average value for 512 samples in the evaluation set.

Table \ref{table:result} demonstrated that 
proposed two-stage learning ({\bf AL+RCCL}) significantly improved the performance
(e.g., about 5~dB in SDR) over the adversarial learning only ({\bf AL}).
Although the experiments were conducted in an anechoic condition,
which is ideal for the proposed method,
the performance of the proposed method is comparable to supervised learning,
which shows the promise of the proposed method.
Figure \ref{fig:result_box} clearly shows that
{\bf AL+RCCL} is much less sensitive to differences in seed values
than {\bf AL} in all metrics.
For example, the distance in SDR between the minimum and maximum was more than 2~dB in {\bf AL}
while that in {\bf AL+RCCL} was about 0.6~dB, which is even lower than that in {\bf PIT} (i.e., 0.9~dB).
These results suggest that the use of remix-cycle-consistent learning effectively increases the accuracy of speech separation and stabilizes adversarial learning.

\section{Conclusion}
\label{sec:conclusion}

This paper proposed a two-stage adversarial/remix-cycle-consistent learning
for accurate and stable unsupervised speech separation.
The proposed algorithm was designed
to explicitly reduce the residual noise and artifacts
in the output of the neural separator.
The experimental results demonstrated that
the proposed algorithm was effective 
in reducing the speech distortions and stabilizing adversarial learning.


\bibliographystyle{IEEEbib}
\small{
\bibliography{refs}
}

\end{document}